\documentclass[%
 aip,
rsi,%
 amsmath,amssymb,
 reprint,%
]{revtex4-1}

\usepackage{graphicx}
\usepackage{dcolumn}
\usepackage{bm}

\usepackage{multirow}
\usepackage{hhline,booktabs,amsmath}
\usepackage{xcolor}

\begin{document}

\preprint{AIP/123-QED}

\title{Spin-glass behavior in Shastry-Sutherland lattice of Tm$_\textbf{2}$Cu$_\textbf{2}$In}

\author{Baidyanath Sahu}
 \altaffiliation{Corresponding author}
 
 \author{R. Djoumessi Fobasso}
 \affiliation{Highly Correlated Matter Research Group, Physics Department, University of Johannesburg, PO Box 524, Auckland Park 2006, South Africa
 }%
 
\author{Andr\'{e} M. Strydom}%
\affiliation{Highly Correlated Matter Research Group, Physics Department, University of Johannesburg, PO Box 524, Auckland Park 2006, South Africa
}%


\begin{abstract}
The spin$\textendash$glass behavior in the ferromagnetic phase of $\mathrm{Tm_2Cu_2In}$ was investigated by dc$\textendash$, ac$\textendash$magnetization, and non-equilibrium dynamics characterizations. An argon arc$\textendash$melted polycrystalline sample of the compound of $\mathrm{Tm_2Cu_2In}$ was found to adopt the $\mathrm{Mo_2FeB_2}$$\textendash$type tetragonal structure (space group $P_4/mbm$). The temperature variation of dc$\textendash$magnetization exhibits a ferromagnetic behavior with Curie temperature $T_C$ = 32.5 K. The zero field cooled (ZFC) and field cooled (FC) magnetic curves show the thermomagnetic irreversible behavior below $T_C$. The field dependence of irreversible temperature follows the Almeida$\textendash$Thouless line. The frequency and ac$\textendash$driven field dependent anomalies in ac$\textendash$susceptibility results indicate the existence of a spin$\textendash$glass state in $\mathrm{Tm_2Cu_2In}$. The time dependence of magnetization further supports the spin$\textendash$glass behavior observed in $\mathrm{Tm_2Cu_2In}$. The frequency dependence of the freezing temperature in the real part of ac$\textendash$susceptibility has been analyzed on the basis of a power$\textendash$law divergence and Vogel$\textendash$Fulcher law. The obtained results reveal that $\mathrm{Tm_2Cu_2In}$ belongs to the ferromagnetic canonical spin$\textendash$glass class of compounds.

%
 \end{abstract}

\maketitle



Ternary rare-earth based RTX (R = rare-earth, T = transition metals, and X = p block elements), magnetic materials have been intensively studied because of their exotic physical properties \cite{RTX,R2TX3,R2T2X,book1}. Among them, the $\mathrm{R_2T_2X}$ series of compounds are of interest due to their diversity of magnetic properties due to the crystal structure. Till date, $\mathrm{R_2T_2X}$ compounds are mainly reported to form into two types of crystal structure, namely tetragonal and orthorhombic \cite{R2T2X,USi2,RE2Cu2In}. The tetragonal structure of $\mathrm{R_2T_2X}$ with space group $P_4/mbm$ is well known as the Shastry$\textendash$Sutherland lattice systems. The arrangement of rare earth ions as triangles in the Shastry$\textendash$Sutherland lattice favours the occurrence of magnetic frustration \cite{Ce2Pd2Sn,Pr2Pd2In,Er2Pd2Sn}. It is reported that the tetragonal structure of $\mathrm{Gd_2Au_2Cd}$ exhibits ferromagnetic spin$\textendash$glass anomalies \cite{Gd2Au2Cd}.  

The geometrically frustrated lattice in a magnetic system can exhibit spin$\textendash$glass behavior. Since last few decades, spin$\textendash$glass studies remain an attractive topic in magnetism \cite{Binder, Mydosh, book}. Tm-based intermetallic compounds have been paid significant attention in the search for low temperature magnetic properties due to their relatively low transition temperature. In this paper, the magnetic properties of $\mathrm{Tm_2Cu_2In}$ have systematically been investigated. The obtained ferromagnetic spin$\textendash$glass behavior is reported.

A polycrystalline sample of $\mathrm{Tm_2Cu_2In}$ was prepared by the standard arc-melting technique on a water-cooled copper hearth under high purity Argon atmosphere. The high-purity ($\geqslant$ 99.9 wt \%) elements with the stoichiometric 2:2:1 ratio was melted. The ingot was flipped and remelted several times to improve the homogeneity and reaction among the constituents. The weight loss of the melted ingots was less than 0.5 wt.\%. The melted ingot of $\mathrm{Tm_2Cu_2In}$ was then wrapped in tantalum foil and was kept in a highly evacuated sealed quartz tube. The sample was annealed at 973 K for a week. The crystal structure of the annealed sample was confirmed from the powder x-ray diffraction (XRD) pattern using a Rigaku Diffractometer with copper K$\alpha$ radiation. The dc$\textendash$magnetization and ac$\textendash$magnetic susceptibility were measured by using a Dynacool$\textendash$Physical Properties Measurement System (PPMS), from Quantum$\textendash$Design, San Diego. 


\begin{figure}[h!]
      \centering
       \includegraphics[scale =0.32]{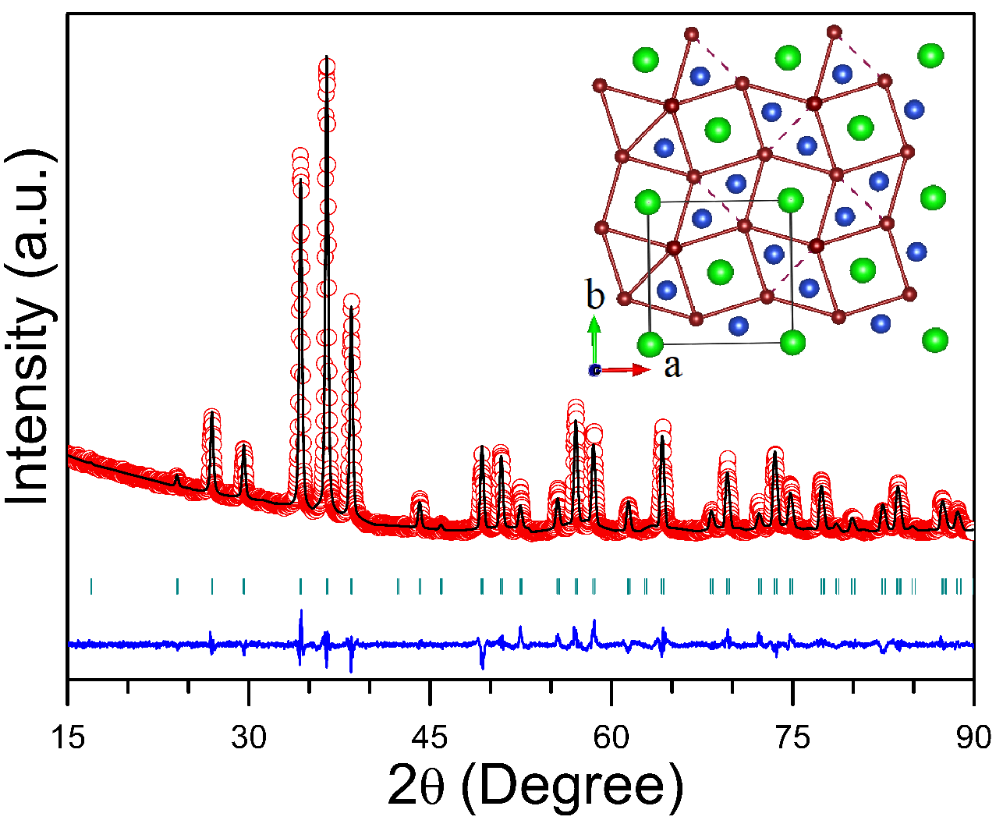}
        \caption{Room temperature powder X$\textendash$ray diffraction patterns along with the refinement profile of $\mathrm{Tm_2Cu_2In}$. Red symbols represent the experimental data and the black line represents the calculated data. The difference between experimental and calculated data is shown as a blue line. A set of vertical bars represent the Bragg peak positions of the tetragonal $\mathrm{Mo_2FeB_2}$ type structure with the space group $P_4/mbm$. Inset: The schematic representation of the tetragonal crystal structure of $\mathrm{Tm_2Cu_2In}$. The wine colour balls are for Tm, blue colour are for Cu and green colour are for In.}
        \label{XRD}
\end{figure}

The powder XRD data at room temperature collected on the polycrystalline sample of $\mathrm{Tm_2Cu_2In}$ were analyzed by Rietveld structural refinement using FULLPROF software \cite{rietveld,fullprof}. Fig.~\ref{XRD} shows the XRD pattern of $\mathrm{Tm_2Cu_2In}$ along with the Rietveld fitted profile. The Rietveld refinement of the XRD data reveals that $\mathrm{Tm_2Cu_2In}$ shows single phase compound of  $\mathrm{Mo_2FeB_2}$$\textendash$type tetragonal crystal structure with space group $P_4/mbm$. The obtained crystallographic parameters from the Rietveld refinement are listed in Table~\ref{refinedata}. The observed crystallography parameters are very close to the values previously reported \cite{USi2,R2Cu2In}. There are two Tm$\textendash$Tm bond distances in the structure. The shortest one measures 0.3656 nm (shown as dotted line) while the longer one measures 0.3864 nm (shown as solid line).

\begin{table}
\caption{\label{comparisrisionMCE} The lattice parameters and unit cell volume of $\mathrm{Tm_2Cu_2In}$ compound obtained from Rietveld refinement of XRD patterns for the tetragonal phase along with the atomic coordinate positions.}
\begin{ruledtabular}
\begin{tabular}{ccccc}
\hspace{-1.5 in} a 	& \hspace{-0.0 in}  & \hspace{-0.5 in} & \hspace{-0.1 in}  & \hspace{-0.7 in}7.399(2)~\AA    \\
\hspace{-1.5 in} b 	& \hspace{-0.0 in}  & \hspace{-0.5 in} & \hspace{-0.1 in}  & \hspace{-0.7 in} 7.399(2)~\AA \\
\hspace{-1.5 in} c  & \hspace{-0.0 in}  & \hspace{-0.5 in} & \hspace{-0.1 in}  & \hspace{-0.7 in} 3.697(2)~\AA \\
\hspace{-1.5 in} V  & \hspace{-0.0 in}  & \hspace{-0.5 in} & \hspace{-0.1 in}  & \hspace{-0.7 in} 202.368(2)~\AA$^3$ \\
\hline
\\
\hspace{0.1 in} Atomic coordinates  for $\mathrm{Tm_2Cu_2In}$ 	& \hspace{-0.2 in}  & \hspace{-0.2 in} & \hspace{-0.2 in}     & \hspace{-0.2 in} \\
\hline
\\
 \hspace{-1.5 in}Atom & \hspace{-2.5 in}Wyckoff & \hspace{-1.5 in}$x$ & \hspace{-0.7 in}$y$ & \hspace{-0.2 in}$z$ \\
 \hline
 & & & & \\
 \hspace{-1.5 in}Tm & \hspace{-2.5 in}$4g$ & \hspace{-1.5 in}0.1747(1) & \hspace{-0.7 in}0.6747(1) & \hspace{-0.2 in}1/2\\
 \hspace{-1.5 in}Cu & \hspace{-2.5 in}$4h$ & \hspace{-1.5 in}0.6212(1) & \hspace{-0.7 in}0.1212(1) & \hspace{-0.2 in}0\\
 \hspace{-1.5 in}In & \hspace{-2.5 in}$2a$ & \hspace{-1.5 in}0 & \hspace{-0.7 in}0 & \hspace{-0.2 in}0\\
\end{tabular}
\end{ruledtabular}
\label{refinedata}
\end{table}


\begin{figure}[h!]
      \centering
       \includegraphics[scale =0.32]{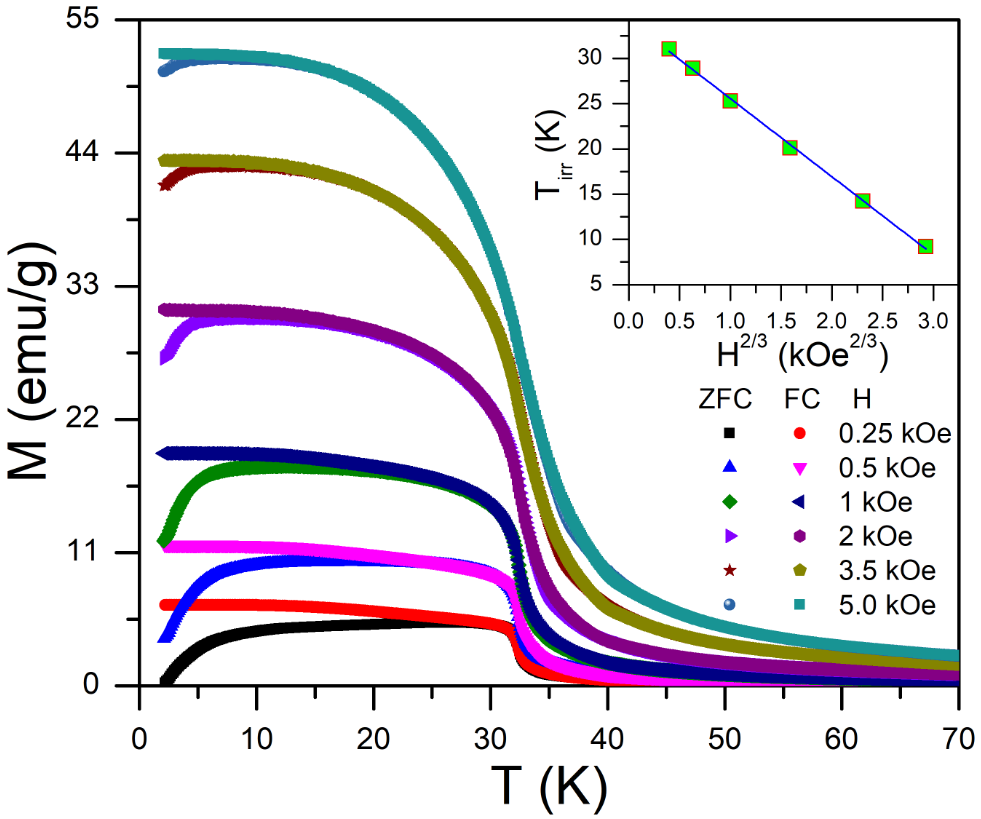}
       \caption{Main panel: Temperature dependence of dc$\textendash$magnetization of $\mathrm{Tm_2Cu_2In}$ under different magnetic field. Inset; $\mathrm{T_{irr}}$ versus $H^{2/3}$ plot following the Almeida$\textendash$Thouless law.}
       \label{MT}
\end{figure}

The zero field cooled (ZFC) and field cooled (FC) dc$\textendash$magnetization as function of temperature for different external applied magnetic fields starting from 0.25 kOe to 5 kOe of $\mathrm{Tm_2Cu_2In}$ are shown in Fig.~\ref{MT}. FC magnetization shows a typical ferromagnetic behavior with transition temperature $T_C$ = 32.5~K. The obtained $T_C$ value is in good agreement with the earlier report \cite{RE2Cu2In, Tm2Cu2In}. The ZFC and FC curves start to diverge below a certain temperature, called the irreversible temperature (${T_{irr}}$). It is seen that ${T_{irr}}$ strongly depends on the applied magnetic field. The experimentally obtained ${T_{irr}}$ for different applied magnetic fields was estimated from the ZFC and FC magnetization. Inset of Fig.~\ref{MT} shows the $H$$\textendash$$T$ phase diagram by plotting of $T_{irr}$ verses $H^{2/3}$. This $H$$\textendash$$T$ phase diagram follows the Almeida$\textendash$Thouless line for strong irreversibility characteristics \cite{AT,GT}. The results of dc$\textendash$magnetization and the Almeida$\textendash$Thouless line have provided indications for the presence of spin$\textendash$glass behavior in the compound.       

\begin{figure}[h!]
      \centering
       \includegraphics[scale =0.325]{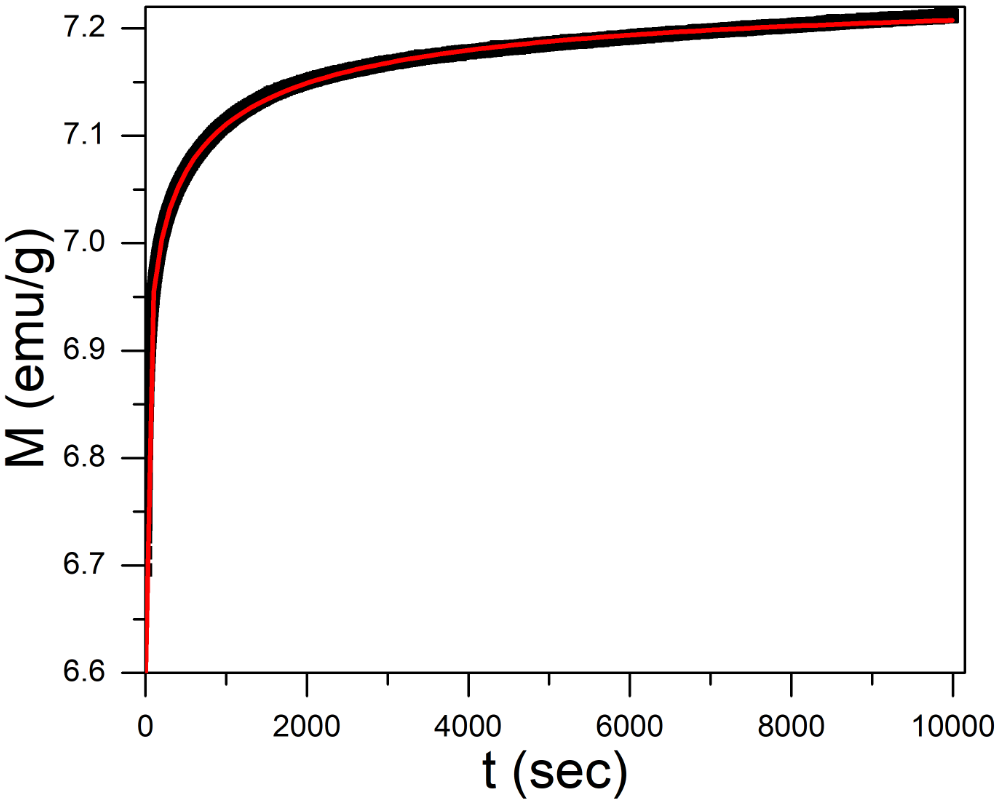}
        \caption{(a): The time dependence of relaxation magnetization of $\mathrm{Tm_2Cu_2In}$ measured at 6 K after zero applied magnetic field and measured under a magnetic field of 500 Oe. The red curve is the fit of the superposition of a stretched exponential expression and a constant term, see Eq.~\ref{mt}.}
       \label{time}
\end{figure}

Spin$\textendash$glass behavior may also be characterized by the magnetic relaxation method. The relaxation process can be measured by both ZFC and FC conditions. In this present case, it is measured under ZFC condition. Here, the sample was cooled without external magnetic field from the paramagnetic region to 6 K (below the freezing temperature). After reaching the desired temperature, a waiting time of 60 sec was observed and the time evolution of the magnetization [M(t)] was recorded under a tiny applied magnetic field. Fig.~\ref{time} shows the time dependence of magnetization $M(t)$ measured at 6 K in ZFC condition under the field value of 500 Oe. As seen in Fig.~\ref{time}, the magnetization increases slowly with time and does not saturate even after a time of 10000 s. The time dependence of the magnetization is explained by the standard stretched exponential function, as follows \cite{mt1, mt2}.

\begin{eqnarray}
M(t) = M_0-M_gexp\left[-\left(\frac{t}{\tau}\right)^{\beta}\right],
\label{mt}
\end{eqnarray}
where $M_0$ is the spontaneous magnetization and $M_g$ is related with the glassy component of magnetization. $\tau$ and $\beta$ are the characteristic parameters for relaxation time constant, and stretching exponent respectively. The LSQ fitted line of Eq.~\ref{mt} to the experimental data in Fig.~\ref{time} yielded the parameters $M_0$ = 7.2 emu/g, $M_g$ = 0.6 emu/g, the relaxation time $\tau$ = 175 sec and $\beta$ = 0.30. For a typical spin$\textendash$glass system, the value of $\beta$ lies between 0 and 1 \cite{mt3,DeltaTf}. As per earlier reports, our result supports that $\mathrm{Tm_2Cu_2In}$ exhibits a spin$\textendash$glass state.

\begin{figure}[h!]
      \centering
       \includegraphics[scale =0.40]{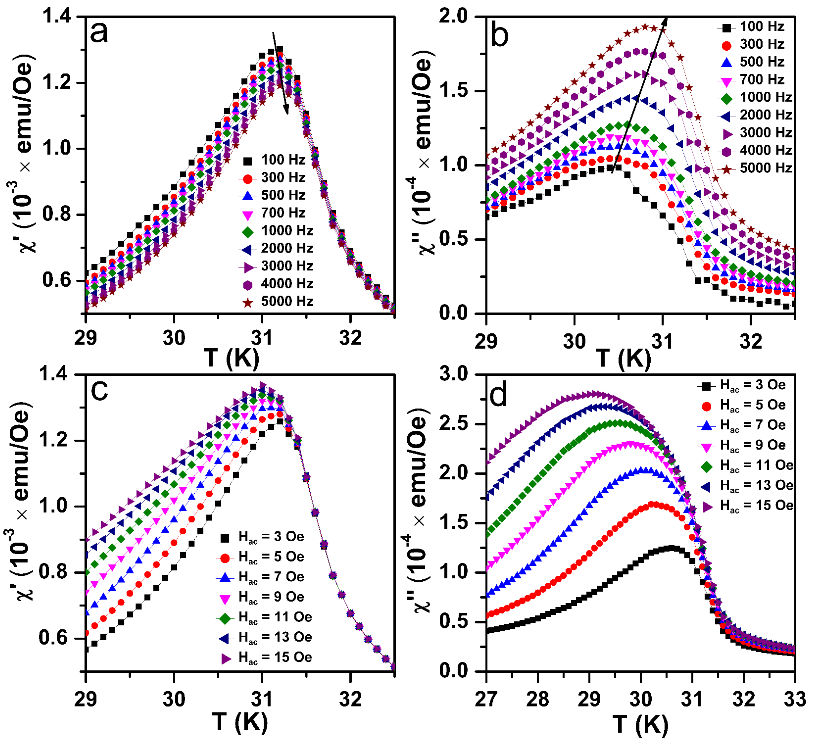}
        \caption{(a): The temperature dependence of the real parts of the ac$\textendash$magnetic susceptibility measured at different frequencies in an applied ac$\textendash$magnetic field of 3 Oe. (b): The temperature dependence of the imaginary parts of the ac magnetic susceptibility measured at different frequencies in an applied ac$\textendash$magnetic field of 3 Oe. (c): The temperature dependence of the real part of the ac$\textendash$magnetic susceptibility measured at different ac$\textendash$driven field in an applied frequency 1000 Hz. (d): The temperature dependence of the imaginary part of the ac$\textendash$magnetic susceptibility measured at different ac$\textendash$driven fields ($H_{ac}$) in an applied frequency of 1000 Hz.}.
       \label{AC}
\end{figure}

The ac$\textendash$magnetic susceptibility is an important characterization to confirm and categorize spin$\textendash$glass systems. Figs.~\ref{AC}(a) and (b) show the temperature dependence of the real ($\chi'(T)$) and imaginary ($\chi''(T)$) parts of the ac$\textendash$magnetic susceptibility for different frequencies, respectively, where the dc$\textendash$magnetic field was zero and ac$\textendash$driven field ($H_{ac}$) = 3 Oe. As seen in Fig.~\ref{AC}(a), $\chi'(T)$ shows a pronounced peak around the spin$\textendash$glass transition temperature. The peak temperature is known as the spin freezing temperature ($T_f$). It is observed that $T_f$ slowly shifts to higher temperature with increasing frequency. Similarly, a frequency dependent peak is also observed in  $\chi''(T)$. The magnitude of the peak of $\chi''(T)$ increases with increasing the frequency, which is related to the dissipation process in the system. These results further confirm the spin$\textendash$glass behavior in the system.

To further explore the spin$\textendash$glass state in $\mathrm{Tm_2Cu_2In}$, temperature dependent ac$\textendash$susceptibility was measured under various ac$\textendash$driven field ($H_{ac}$). Fig.~\ref{AC}(c) and (d) show the $\chi'(T)$ and $\chi''(T)$ of the ac$\textendash$magnetic susceptibility for different $H_{ac}$ and at a constant frequency of 1000 Hz with zero dc$\textendash$magnetic field respectively. As shown in Fig.~\ref{AC}(c) and (d), the position of the peak in $\chi'(T)$ and $\chi''(T)$ depend on $H_{ac}$. Both $\chi'(T)$ and $\chi''(T)$ shifts to low temperature and the magnitude increases with increasing $H_{ac}$. These observations also confirm the ferromagnetic spin$\textendash$glass behavior in $\mathrm{Tm_2Cu_2In}$.

\begin{figure}[h!]
      \centering
       \includegraphics[scale =0.475]{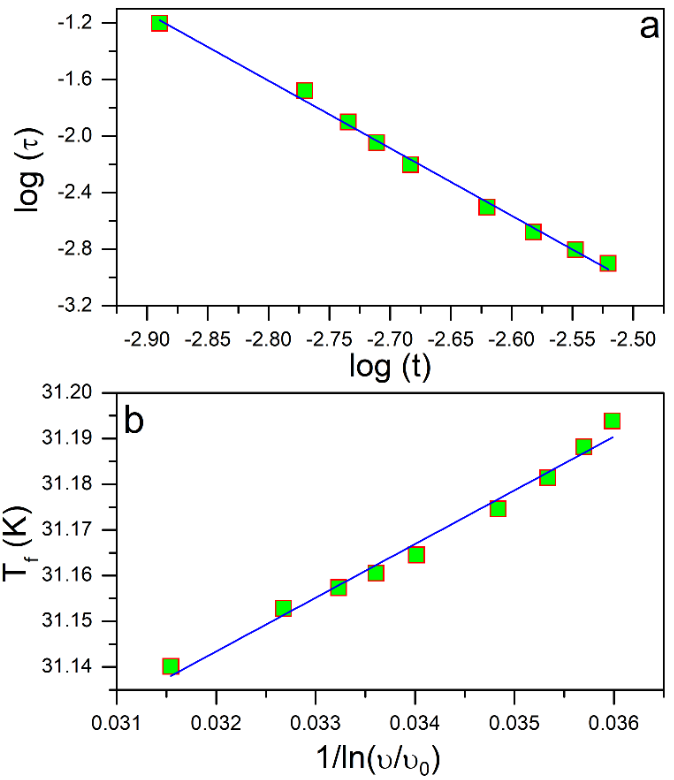}
        \caption{(a): log($\tau$) $vs.$ log(t) plot for the frequency dependence of freezing temperature, where the reduced temperature t = ($T_f$ $-$ $T_{SG}$)/$T_{SG}$. The solid line represents the fitting of power-law divergence to experimental data. (b) The frequency dependence of freezing temperature plotted as $T_f$ $vs.$ ln($\mathrm{\nu_0}$/$\mathrm{\nu}$). The solid straight line represents the fitting of Vogel-Fulcher law to experimental data.}.
       \label{VF}
\end{figure}

The ac$\textendash$magnetic susceptibility results confirm that $\mathrm{Tm_2Cu_2In}$ follows spin$\textendash$glass type transition with a freezing temperature $T_f$~$\approx$~31.1 K. In a typical spin$\textendash$glass system, the frequency dependence of $T_f$ is used to compare the relative shift in freezing temperature per decade of frequency by the following expression \cite{DeltaTf}:
\begin{equation}
\delta {T_f} = \frac{\Delta {T_f}}{[{T_f} \Delta \log_{10} {\nu}]}.
\end{equation}

Here, $\Delta T_{f}$ = $T_{f2}~-~T_{f1}$, and $\Delta$log$_{10}$$\nu$ = log$_{10}$$\nu_2$~-~log$_{10}$$\nu_1$ with 2$\pi\nu_1$ = 100 Hz and 2$\pi\nu_2$ = 1000 Hz. It is found a value of $\delta T_f$ = 0.0008 for $\mathrm{Tm_2Cu_2In}$, which lies typically in the range of a canonical spin glass system \cite{DeltaTf}. This result revealed that a suitable characterization of $\mathrm{Tm_2Cu_2In}$ is based on the canonical spin$\textendash$glass state.

The frequency dependence of $T_f$ for $\mathrm{Tm_2Cu_2In}$ follows the conventional power$\textendash$law divergence of a critical slowing down and is given by the relation \cite{Binder,Mydosh,TF};

\begin{equation}
\tau = \tau_0 \left(\dfrac{T_f-T_{SG}}{T_{SG}}\right) ^ {z\nu'} ,
\label{power-law}
\end{equation}
where $\tau = 1/\nu$ is the relaxation time corresponding to the measured frequency ($f = 2\pi\nu$) and $\tau_0$ is a characteristic relaxation time of single spin flip. $T_{SG}$ is known as the spin$\textendash$glass temperature as frequency tends to zero. $\nu'$ is the critical exponent of correlation length, $\xi$ = (($T_f$/$T_{SG}$)-1)$^{\nu'}$ and $\tau$ = $\xi^z$. $z\nu'$ is known as the dynamic critical exponent and is an important parameter of the spin$\textendash$glass system. The critical exponent $z\nu'$ lies between 4 and 12 for a typical spin$\textendash$glass system. Eq.~\ref{power-law} can be written as:
\begin{equation}
\mathrm{log}(\tau) = \mathrm{log}(\tau_0) - {z\nu'}~\mathrm{log}(t) ,
\end{equation}
where $t$ = $\left(\dfrac{T_f-T_{SG}}{T_{SG}}\right)$. The frequency dependence of $T_F$ obtained from $\chi'(T)$ is plotted as log$\textendash$log of inverse frequency $\tau$ against $t$ and is shown in Fig.~\ref{VF}(a). The intercept and slope of the linear fit to log($\tau$)$\textendash$log($t$) are used to calculate the value of $\tau_0$ and $z\nu'$ respectively. From the best fit, $\tau_0$ = 1.04 $\times$10$^{-15}$ s and $z\nu'$ = 4.8 are obtained. The value of $z\nu'$ supports, the observation of spin$\textendash$glass system in $\mathrm{Tm_2Cu_2In}$. The obtained value of $\tau_0$ for $\mathrm{Tm_2Cu_2In}$ is close to the typical value of 10$^{-12}$~$\textendash$~10$^{-13}$ s for canonical spin$\textendash$glass systems. 

A spin$\textendash$glass system is also categorised using the dynamical Vogel-Fulcher scaling law~\cite{Mydosh,VF}. The frequency dependence of $T_f$ from $\chi'(T)$ is described by:
\begin{equation}
\nu = \nu_0 \exp \left[ \frac{-\mathrm{E_a}}{\mathrm{k_B(T-T_0)}}\right] ;
\label{vogel}
\end{equation}
where where $\mathrm{k_B}$ stand for the Boltzmann constant, $\nu_0$ = 1/$\tau_0$ is the characteristic attempt frequency, E${_a}$ is thermal activation energy, and $T_0$ is known as the Vogel-Fulcher temperature. To conveniently fit the data, Eq.~\ref{vogel} can be rewritten as:
\begin{equation}
T_f = T_0+\dfrac{E_a/k_B}{\mathrm{ln}(\nu_0/\nu)}. 
\label{eq}
\end{equation}
According to this equation, a plot of $T_f$ $vs.$ 1/ln$(\nu_0/\nu)$ is made. The slope and intercept of the linear fit are used to estimate ${E_a/k_B}$ and $\tau_0$ respectively. The best fit yielded a value of $E_a/k_B$ = 11.75 K and $T_0$ = 30.77 K. The ratio of ($E_a/k_B$)/$T_0$ is less than 1 confirming the system is in canonical spin$\textendash$glass behavior. However, ($E_a/k_B$)/$T_0$ is generally found to be relatively large in cluster spin$\textendash$glass state. This result revealed that $\mathrm{Tm_2Cu_2In}$ exhibits ferromagnetic canonical spin$\textendash$glass state for the geometrical frustration in the crystal structure.


\subsection{\label{sec:level2}Summary}

In summary, we have successfully prepared a polycrystalline sample of the intermetallic compound $\mathrm{Tm_2Cu_2In}$, which crystallize in the $\mathrm{Mo_2B_2Fe}$~$\textendash$~type of tetragonal structure. The spin$\textendash$glass behavior in the ferromagnet found below its characteristic freezing temperature $T_f$ = 31.1 K, has been confirmed by dc$\textendash$magnetization and ac$\textendash$susceptibility measurements. The spin$\textendash$glass behavior in $\mathrm{Tm_2Cu_2In}$ has also been suggested by the magnetic relaxation behavior. The canonical spin$\textendash$glass state in $\mathrm{Tm_2Cu_2In}$ was suggested from the experimental data of frequency dependence of $T_f$ in ac$\textendash$susceptibility. The experiment results are discussed with the theoretical model of power divergence method and Vogel$\textendash$Fulcher law. This observation revealed that the investigated compound exhibits ferromagnetic canonical spin$\textendash$glass state due to the geometrically frustration in the structure likely due to the formation of distorted triangles by the magnetic rare-earth ions.

%
%
%

\subsection*{\label{sec:level3}Acknowledgements}   
This work is supported by Global Excellence and Stature (UJ-GES) fellowship, University of Johannesburg, South Africa. DFR thanks OWSD and SIDA for the fellowship towards PhD studies. AMS thanks the URC/FRC of UJ for assistance of financial support.

\subsection*{\label{sec:level3}DATA AVAILABILITY}
The data that support the findings of this study are available
from the corresponding author upon reasonable request.


\subsection*{\label{ref}References}


\begin{thebibliography}{}




\bibitem{RTX} S. Gupta, and  K.G. Suresh,  J. Alloys Compd., \textbf{618} (2015) 562$\textendash$606.

\bibitem{R2T2X} Y. Zhang, J. Alloys Compd., \textbf{787} (2019) 1173$\textendash$1186.

\bibitem{R2TX3} P. Zhi-Yan, C. Chong-De, B. Xiao-Jun, S. Rui-Bo, Z. Jian-Bang, and D. Li-Bing, Chin. Phys. B, \textbf{22} (2013) 056102.

\bibitem{book1} H. S. Li, and J. M. D. Coey, "Magnetic properties of ternary rare-earth transition-metal compounds", Handbook of magnetic materials \textbf{6} (1991) 1$\textendash$83.

\bibitem{USi2} M. Y. Lukachuk, and R. Pöttgen, Z. Kristallogr, Cryst. Mater., \textbf{218} (2003) 767$\textendash$787.

\bibitem{RE2Cu2In} I. R. Fisher, Z. Islam, and P. C. Canfield, J. Magn. Magn. Mater \textbf{202} (1999) 1-10.

\bibitem{Ce2Pd2Sn} J. G. Sereni, M. G. Berisso, G. Schmerber, and J. P. Kappler, Phys. Rev. B \textbf{81} (2010) 184429.


\bibitem{Pr2Pd2In} P. Fischer, T. Herrmannsd\"{o}rfer, T. Bonelli, F. Fauth, L. Keller, E. Bauer, and M. Giovannini, J. Phys. Condens. Matter \textbf{12} (2000) 7089.

\bibitem{Er2Pd2Sn} G. Saucedo Salas, S. Baidyanath, A. Strydom, S. Calder, and H. Nair, Bulletin of the American Physical Society (2020).


\bibitem{Gd2Au2Cd} S. Rayaprol, and R. P\"{o}ttgen, Phys. Rev. B \textbf{73} (2006) 214403.


\bibitem{Binder}K. Binder and A.P. Young, Rev. Mod. Phys. \textbf{58} (1986) 801.
	
	
\bibitem{Mydosh}J. A. Mydosh, Spin Glasses (Taylor and Francis, London), 1993.

\bibitem{book}Spin Glasses and Random Fields, edited by A. P. Young (World Scientific, Singapore), 1997.

\bibitem{rietveld} H. M. Rietveld, J. Appl. Crystallogr., \textbf{2} (1969) 65.

		
\bibitem{fullprof} J. Rodriguez-Carvajal, Fullprof Suite http://www. ill. eu/sites/fullprof/ (2017).



\bibitem{R2Cu2In} Y. M. Kalychak, V.I. Zaremba, V.M. Baranyak, P. Y. Zavalii, V. A. Bruskov, L. V. Sysa L.V., and O. V. Dmytrakh, Izv. Akad. Nauk SSSR, Neorg. Mater., \textbf{26} (1990) 74$\textendash$76. 


\bibitem{Tm2Cu2In} Y. Zhang, Y. Yang, X. Xu, L. Hou, Z. Ren, X. Li, and G. Wilde, J. Phys. D: Appl. Phys. \textbf{49} (2016) 145002.



\bibitem{AT} J. R. L. de Almeida and D. J. Thouless, J. Phys. A \textbf{11} (1978) 983.
	
\bibitem{GT} M. Gabay and G. Toulouse, Phys. Rev. Lett. \textbf{47} (1981) 201.



\bibitem{mt1} S. Ghara, B. G. Jeon, K. Yoo, K. H. Kim, and A. Sundaresan, Phys. Rev. B \textbf{90} 024413 (2014).


\bibitem{mt2} A. Bhattacharyya, S. Giri, and S. Majumdar, Phys. Rev. B \textbf{83} 134427 (2011).

\bibitem{mt3} D. Chu, G. G. Kenning, and R. Orbach, Phys. Rev. Lett. \textbf{72} 3270 (1994).

\bibitem{DeltaTf} J. A. Mydosh, Spin Glasses: An Experimental Introduction
(Taylor and Francis, London, 1993), Chap. 3.

\bibitem{TF} P. C. Hohenberg and B. I. Halperin, Rev. Mod. Phys. \textbf{49}  (1977) 435.



\bibitem{VF}J. Souletie and J. L. Tholence, Phys. Rev. B \textbf{32} (1985) 516.

	
	
	
\end{thebibliography}

\end{document}